%% file: sn-article.tex
\documentclass[pdflatex,sn-mathphys-num]{sn-jnl}

\usepackage{graphicx}%
\usepackage{multirow}%
\usepackage{amsmath,amssymb,amsfonts}%
\usepackage{amsthm}%
\usepackage{mathrsfs}%
\usepackage[title]{appendix}%
\usepackage{xcolor}%
\usepackage{textcomp}%
\usepackage{manyfoot}%
\usepackage{booktabs}%
\usepackage{algorithm}%
\usepackage{algorithmicx}%
\usepackage{algpseudocode}%
\usepackage{listings}%
\usepackage{multirow}
\usepackage{svg}
\usepackage{tabularx}

\begin{document}
	
	\title[Real-Time Cloth Simulation Using WebGPU: Evaluating Limits of High-Resolution]{Real-Time Cloth Simulation Using WebGPU: Evaluating Limits of High-Resolution Cloth Model}
		
	\author[1]{\fnm{Nak-Jun} \sur{Sung}}\email{njsung@sch.ac.kr}
	\author[1]{\fnm{Jun} \sur{Ma}}\email{majun@sch.ac.kr}
	\author[1]{\fnm{TaeHeon} \sur{Kim}}\email{kemca3249@sch.ac.kr}
	\author[2]{\fnm{Yoo-joo} \sur{Choi}}\email{yjchoi@smit.ac.kr}
	\author[2]{\fnm{Min-Hyung} \sur{Choi}}\email{min.choi@slu.edu}
	\author*[4]{\fnm{Min} \sur{Hong}}\email{mhong@sch.ac.kr}

	\affil[1]{\orgdiv{Dept of Software Convergence}, \orgname{Soonchunhyang University}, \orgaddress{\street{Soonchunhyang-ro 22}, \city{Asan-si}, \postcode{31538}, \state{Chungcheongnam-do}, \country{South Korea}}}
	
	\affil[2]{\orgdiv{Department of AI Software Engineering}, \orgname{Seoul Media Institute Technology}, \orgaddress{\street{99, Hwagok-ro 61-gil}, \city{Seoul-si}, \postcode{07590},\country{South Korea}}}
	
	\affil[3]{\orgdiv{Department of Computer Science}, \orgname{Saint Louis University}, \orgaddress{\street{220 North Grand Blvd}, \city{Saint Louis}, \postcode{63103}, \state{Missouri}, \country{USA}}}
	
	\affil*[4]{\orgdiv{Dept of Computer Software Engineering}, \orgname{Soonchunhyang University}, \orgaddress{\street{Soonchunhyang-ro 22}, \city{Asan-si}, \postcode{31538}, \state{Chungcheongnam-do}, \country{South Korea}}}
	
	\abstract{This study explores the capabilities of WebGPU, an emerging web graphics paradigm, for real-time cloth simulation. Traditional WebGL-based methods have been in handling complex physical simulations due to their emphasis on graphics rendering rather than general-purpose GPU (GPGPU) operations. WebGPU, designed to provide modern 3D graphics and computational capabilities, offers significant improvements through parallel processing and support for computational shaders. In this work, we implemented a cloth simulation system using the Mass-Spring Method within the WebGPU framework, integrating collision detection and response handling with the 3D surface model. First, comparative performance evaluations demonstrate that WebGPU substantially outperforms WebGL, particularly in high-resolution simulations, maintaining 60 frames per second (fps) even with up to 640K nodes. The second experiment aimed to determine the real-time limitations of WebGPU and confirmed that WebGPU can handle real- time collisions between  4K and 100k cloth node models and a 100K triangle surface model in real-time. These experiments also highlight the importance of balancing real-time performance with realistic rendering when handling collisions between cloth models and complex 3D objects. Our source code is available at \href{https://github.com/nakjun/Cloth-Simulation-WebGPU}{https://github.com/nakjun/Cloth-Simulation-WebGPU}.
    }
	
	
	\maketitle
	
	\input{introduction.tex}
	\input{related_works.tex}
	\input{method.tex}

	\input{experiments.tex}

\input{conclusion.tex}
	
	\bmhead{Acknowledgements}
	
	This research was supported by the Basic Science Research Program through the National Research Foundation of Korea (NRF) funded by the Ministry of Education under Grant NRF-2022R1I1A3069371, was funded by the BK21 FOUR (Fostering Outstanding Universities for Research) No. 5199990914048.
	
	\newpage

\input{reference}
    
    \newpage
    \input{appendix}
	
\end{document}

%% file: introduction.tex
\section{Introduction} 

Real-time cloth simulation is essential in various fields, including computer graphics~\cite{lan2024efficient}, virtual reality~\cite{va2021real}, gaming~\cite{tang2016cross,kim2012long}, and virtual fitting~\cite{dai2024fabric,chen2024dress,wu2022gpu}, where interactive and visually realistic representations of fabrics are required. To achieve real-time performance in these simulations, efficient physics-based computation and rendering techniques are necessary to enable smooth and responsive simulation performance. This optimization is particularly crucial in web environments, where simulations must be accessible to users across different platforms and devices without the installation of specialist software. Harnessing the power of web graphics technologies ensures that these advanced simulations are widely available and easily accessible, expanding their potential impact and user base. As the web continues to be the primary medium for delivering interactive content, developing real-time simulation techniques in web environments is becoming increasingly important, driving innovation and improving user experiences across diverse applications.

WebGL is a widely adopted JavaScript API that enables 3D graphics rendering in web browsers without the need for external plugins, building on the foundations of OpenGL ES 2.0. This technology has significantly contributed to the widespread adoption of complex and interactive 3D graphics on the web, allowing developers to create immersive experiences accessible across multiple devices. However, WebGL has inherent limitations that affect its ability to handle general- purpose GPU (GPGPU) computing tasks, particularly those required for complex simulations and large-scale data processing. Because WebGL is primarily optimized for rendering and does not natively support compute shaders, applications that rely on intensive physics simulations may experience performance bottlenecks. Furthermore, WebGL’s synchronous API approach restricts asynchronous processing, which can degrade performance in data-intensive or computationally complex scenarios~\cite{li2021ncloth,rzepka2020cloth}. These limitations highlight the need for more advanced web graphics APIs that can efficiently support modern computational and simulation demands.

To address WebGL limitations, the web community has developed a new standard called WebGPU. WebGPU is a modern web standard and JavaScript API designed to provide advanced 3D graphics and compute capabilities in web browsers, offering accelerated graphics processing and support for GPGPU (General-Purpose computing on Graphics Processing Units~\cite{kenwright2022introduction}. WebGPU development began in 2017 through a collaboration of the W3C GPU for the Web Community Group and engineers from major technology companies such as Apple, Mozilla, Microsoft, and Google. Official support for WebGPU started with Chrome version 113, and as of 2024, it is integrated into most major browsers, including Firefox and Safari. Unlike WebGL, which is primarily focused on rendering, WebGPU supports compute shaders and provides a lower-level, more flexible API that can directly leverage modern GPU architectures. This enables more efficient parallel processing and allows developers to implement complex simulations~\cite{gpgpuComputeshader}, physics computations, and large-scale data processing tasks~\cite{tiis_pointcloud} directly in the browser. With these enhanced capabilities, WebGPU addresses the performance bottlenecks inherent to WebGL, paving the way for high-performance, real-time simulations, and interactive 3D graphics applications on the web.

In this paper, we present a real-time cloth simulation system implemented using WebGPU and based on the Mass-Spring Method. We conduct a comprehensive per- formance evaluation to analyze how effectively this system performs compared to simulations built using a WebGL library like ThreeJS. Specifically, we assess the sys- tem’s ability to maintain real-time performance under various scenarios, examining its ability to handle increasing levels of complexity in both the cloth model’s resolution and the 3D surface objects it interacts with. Through these experiments, we demonstrate that complex, high-performance simulations are now achievable in web environments with WebGPU, highlighting the benefits of supporting computational shaders and efficient parallel processing. Our findings aim to provide insights and guidance for researchers and developers interested in building next-generation web-based graphics applications and simulations using WebGPU, taking advantage of its advanced capabilities in accelerated graphics and computation. We therefore release our source code at \href{https://github.com/nakjun/Cloth-Simulation-WebGPU}{https://github.com/nakjun/Cloth-Simulation-WebGPU}. As web graphics environ- ments continue to evolve, the integration of real-time simulations using technologies like our study is expected to enable a wide range of applications across diverse domains, further expanding the impact and usefulness of interactive web-based 3D graphics.

%% file: related_works.tex
\section{related works}
\subsection{WebGPU}
\begin{figure}[H]
	\centering
	\includegraphics[width=1.0\linewidth]{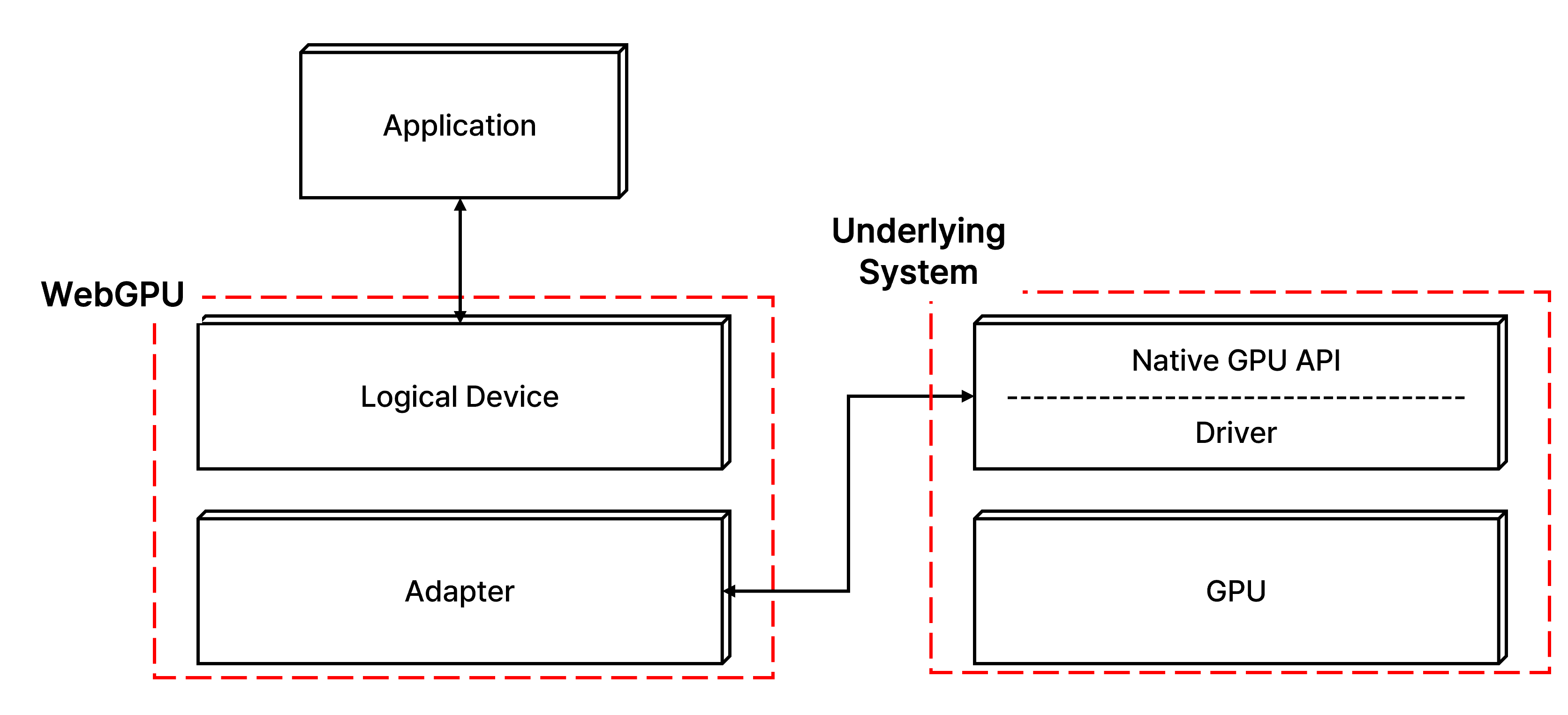}
	\caption{API architecture of the WebGPU}
    \label{fig:API architecture of the WebGPU}
\end{figure}

WebGL, a web graphics API based on OpenGL ES, has been used to render graphical environments in browsers since 2011. However, in order to develop a GPU environ- ment and provide various high-performance services in the web environment, a new web graphics paradigm called WebGPU has emerged. WebGPU focuses less on simple rendering execution and more on sharing GPU resources for general computation. WebGPU includes vertex shader and fragment shader, which are shader pipelines for rendering, similar to WebGL~\cite{kenwright2022introduction}\cite{fransson2023performance}\cite{chickerur2024webgl}. However, it differs from WebGL in that it includes compute shader, which allows GPU resources to be used for general calculations. Figure \ref{fig:API architecture of the WebGPU} represents the architecture of WebGPU API.

By using this WebGPU, simulation systems can be effectively improved, where it was difficult to guarantee real-time performance in existing WebGL. For these reasons, in this study, we implemented a simulation system based on WebGPU.

There has not been much research on this topic yet, as it has only recently been registered as a regular graphics pipeline. Looking at recent research trends, Ziya Usta’s research~\cite{webgpuGIS} analyzed how much performance improvement WebGPU can provide compared to WebGL for GIS systems. Next, Benedikt Peter’s research~\cite{peter2023particle} revealed the details of implementing a particle system using WebGPU. Finally, Fransson et al.’s research~\cite{fransson2023performance} shows that the Godot Engine for games also performs faster in terms of average CPU and GPU fps compared to WebGL environment. It can be seen that most of the research is conducted by comparing with WebGL, an existing web graphics pipeline. Additionally, looking at the research results, we believe that using WebGPU can effectively improve simulation systems where it was difficult to guarantee real-time performance in existing WebGL. For these reasons, in this paper, we implement a WebGPU-based simulation system.

\subsection{Cloth Simulation Method}
There are various techniques for simulating cloth based on basic simulation theory~\cite{baraff1998large}. Terzopoulos et al.’s research~\cite{terzopoulos1987elastically} established foundational principles for physically-based deformable models in computer graphics. Their work introduced elastically deformable models governed by differential equations that realistically simulate dynamic behaviors of flexible materials such as cloth, rubber, and paper. These models effectively capture natural responses to external forces, constraints, and interactions with surrounding environments, enabling realistic simulations of flexible objects in computer graphics. Louchet et al.’s research~\cite{louchet1995evolutionary} proposed an evolutionary approach to identifying parameters for cloth animation models. Specifically, they introduced a mass-spring mesh to model nonlinear elastic behaviors of cloth materials, optimizing internal parameters through evolutionary algorithms. By minimizing the difference between simulated and real cloth behaviors, this method significantly advanced realistic cloth simulations, accurately capturing complex anisotropic stretching and bending properties. Representative simulation methods include the Mass-Spring System(MSS)~\cite{provot1995deformation,liu2013fast}, Position-Based Dynamics(PBD)~\cite{muller2007position,saillant2024high}, Extended Position-Based Dynamics(XPBD)~\cite{macklin2016xpbd}, and Finite Element Method(FEM)~\cite{tan1999constrained,etzmuss2003fast}, and Additionaly, methods derived from these techniques, such as Projective Dynamics(PD)~\cite{bouaziz2014projective} and Vertex Block Descent(VBD)~\cite{chen2024vertex}, are also being studied extensively recently. In addition to these simulation techniques, research on cloth simulation and garment represent using machine learning and deep learning technologies~\cite{chen2024sacanet} is also being actively conducted. 

In this section, we analyze the simulation techniques above to identify the most suitable method for real-time simulation. Each method presents trade-offs between computational complexity, accuracy, and implementation ease: \textbf{MSS} models cloth as interconnected masses and springs, offering simplicity and computational efficiency, but struggles with complex behaviors and numerical stability. \textbf{PBD} directly updates positions using constraints, achieving stable, fast, and easily controlled simulations, though it sacrifices physical accuracy. \textbf{XPBD}, an enhanced version of PBD, provides improved accuracy and stability but at higher computational costs, potentially limiting real-time use. \textbf{FEM} divides cloth into finite elements to calculate stress and strain, producing highly accurate simulations but with significant computational demands. \textbf{PD} projects simulation states onto constraint sets, offering stability and efficiency even at large time steps but requiring careful tuning to achieve realistic behaviors. Lastly, \textbf{VBD} partitions meshes into blocks solved independently, enabling efficient parallel computation but posing challenges in implementation complexity and accuracy. Table \ref{tab:cloth-simulation-criteria} shows the comparison of the above cloth simulation methods.

\begin{table}
	\centering
	\caption{Comparison of cloth simulation methods based on real-time performance and accuracy.}
	\label{tab:cloth-simulation-criteria}
	\begin{tabular}{lcc}
		\toprule
		\textbf{Method} & \textbf{Real-Time Performance} & \textbf{Accuracy} \\
		\midrule
		Mass-Spring System (MSS) & High & Low \\
		Position-Based Dynamics (PBD) & High & Low \\
		Extended Position-Based Dynamics (XPBD) & Medium & Medium \\
		Finite Element Method (FEM) & Low & High \\
		Projective Dynamics & Medium & Medium \\
		Vertex Block Descent & High & Medium \\
		\bottomrule
	\end{tabular}
\end{table}

Many researchers select an appropriate simulation method based on specific requirements regarding performance, accuracy, and complexity. For applications where real-time performance is crucial, such as interactive web simulations or video games, simpler and more efficient methods like MSS or PBD are often chosen, even though they may trade off some accuracy. More advanced methods like XPBD or FEM are preferred when higher accuracy is needed and computational resources allow for more complex calculations.

We focus on implementing a real-time cloth simulation system using MSS, a widely researched method, and WebGPU, a current-trendy web graphics framework. MSS has been chosen for its simplicity, suitability, efficiency for real-time applications, while WebGPU provides the necessary computational capabilities to handle physics calculations efficiently. Additionally, we aim to evaluate and validate the system’s ability to maintain real-time performance by integrating this simulation method into a WebGPU environment. This approach addresses a gap in current research regarding complex real-time simulations in web graphics environment and provides insights and knowhow into the capabilities of WebGPU when implementing commonly used simulation techniques like the MSS. Through this work, we hope to demonstrate that advanced, high-performance cloth simulations can be effectively achieved in modern web environments, thereby contributing to future research and development in web-based graphics and interactive simulations.

%% file: method.tex
\section{Methodology}

\begin{figure}[H]
	\centering
	\includegraphics[width=1.0\linewidth]{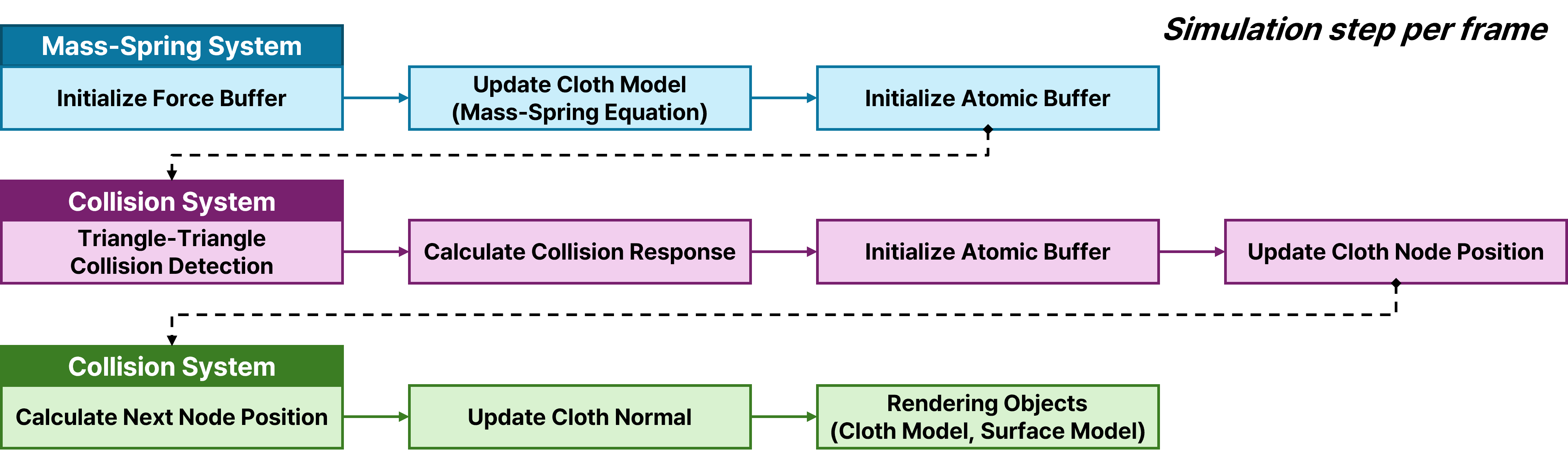}
	\caption{Simulation Steps per Each Frame in our Proposed Simulation System.}
    \label{fig:system overview}
\end{figure}

The proposed simulation system utilizes the MSS architecture and includes a collision detection and response algorithm with a surface model~\cite{baraff1998large}. Our experiment framework consists of the following parts, each performing a specific role: In the MSS, we calculate the spring forces applied to the cloth model based on the relations between nodes and springs, and update the state accordingly. In the collision detection system, we perform narrow phase using triangle-triangle intersection method, which extends the M¨oller–Trumbore intersection algorithm~\cite{moller1997fast}\cite{moller2005fast} between the updated position of the cloth model and the 3d surface model. After detection, we calculate the collision response direction of each triangle and update its post-collision position. In the rendering system, we predict the next position using the Euler method~\cite{lei2012variable}, considering the calculated spring force, external force, gravity, etc. This step is followed by normal vector update and final rendering on the screen. Figure \ref{fig:system overview} shows the proposed simulation pipeline.

The proposed cloth simulation implementation approach follows a spring-centric algo rithm that utilizes atomic functions and a temporary buffer system to accumulate data after computing spring forces. In a node-centric approach, each thread identifies the springs connected to its corresponding node and performs the necessary computations. Unlike the spring-centric approach, this method eliminates the need for atomic functions or a temporary buffer system. However, when extending the simulation to a 3D volumetric object, the node-centric approach becomes infeasible. Therefore, we adopt the spring-centric approach. The detailed algorithm is described in  \textit{Algorithm \ref{alg:Spring-centric Force Calculation Algorithm}} in the Appendix section.

In the collision system, we implement both detection and response to collisions. The system employs a triangle-triangle intersection, an extended M¨oller–Trumbore intersection algorithm, for more detailed collision detection. After detecting the collision, we identify the nodes affected by the collision between triangles and calculate the direction according to the relationship with the collision point. When collision check for all triangle pairs is completed, the node position is updated using the average value of the direction. The detailed collision response algorithm is described in \textit{Algorithm \ref{alg:Collision Response and Position Update Algorithm}} in the Appendix section. The following paragraph presents our proposed extended Möller–Trumbore process.

\paragraph{Extended Möller–Trumbore Process:} Given an edge with endpoints \(\mathbf{S}\) and \(\mathbf{E}\) and a triangle with vertices \(\mathbf{V}_0\), \(\mathbf{V}_1\), \(\mathbf{V}_2\), the algorithm proceeds as follows:
\begin{enumerate}
    \item Compute the edge vector \(\mathbf{d} = \mathbf{E} - \mathbf{S}\) and normalize it to obtain \(\mathbf{r} = \mathbf{d}/\|\mathbf{d}\|\).
    \item Define the triangle edges: \(\mathbf{e}_1 = \mathbf{V}_1 - \mathbf{V}_0\) and \(\mathbf{e}_2 = \mathbf{V}_2 - \mathbf{V}_0\).
    \item Calculate the auxiliary vector \(\mathbf{h} = \mathbf{r} \times \mathbf{e}_2\) and scalar \(a = \mathbf{e}_1 \cdot \mathbf{h}\). If \(|a| < \epsilon\), the edge is parallel to the triangle plane and no intersection occurs.
    \item Otherwise, compute \(f = \frac{1}{a}\), the vector \(\mathbf{s} = \mathbf{S} - \mathbf{V}_0\), and the barycentric coordinate \(u = f\,(\mathbf{s} \cdot \mathbf{h})\). If \(u \notin [0,1]\), the intersection is outside the triangle.
    \item Next, determine \(\mathbf{q} = \mathbf{s} \times \mathbf{e}_1\), \(v = f\,(\mathbf{r} \cdot \mathbf{q})\), and the parameter \(t = f\,(\mathbf{e}_2 \cdot \mathbf{q})\). If \(v < 0\) or \(u + v > 1\) or if \(t\) does not lie within the edge bounds (i.e., \(\epsilon < t < \|\mathbf{d}\|\)), then no valid intersection exists.
    \item If all the above conditions are satisfied, then the intersection point is given by \(\mathbf{P}_{\text{int}} = \mathbf{S} + t\,\mathbf{r}\), which indicates the precise location along the edge where the collision occurs.    
\end{enumerate}

%% file: experiments.tex
\section{Experiments}
We leverage this MSS and collision system to simulate the cloth model and handle collision detection and compute responses when interacting with other objects. Using our proposed simulation system, we conducted performance comparison experiments with simulation systems in the WebGL environment and experiments to find the limits of real-time simulation in the WebGPU environment. Table \ref{tab:hardware-software-environment} lists the hardware and software environments in which the experiment was performed. Since web graphics performance varies depending on the performance of the computer running the web browser, all experiments were conducted under the same conditions.

\begin{table*}[t]
	\centering
	\caption{Hardware and software environment in our research.}
	\label{tab:hardware-software-environment}
	\begin{tabularx}{\textwidth}{Xl}
		\toprule
		\textbf{Name} & \textbf{Spec} \\
		\midrule
		CPU & Intel i7-7700, 3.6GHz \\
		GPU & Nvidia GeForce RTX 4070 Ti, 12GB VRAM \\
		Memory & 32GB \\
		Web Browser & Chrome v122.0.6261.94 \\
		Libraries & webgpu/types 0.1.40, gl-matrix 3.4.3, three : 0.160.1 \\
		Programming Language, frameworks & Typescript, React.js \\
		IDE & Visual studio code \\
		\bottomrule
	\end{tabularx}
\end{table*}

\subsection{Performance Comparison between WebGL and WebGPU Evnrionment: Simulate Hanging Cloth}
\begin{figure}[H]
	\centering
	\includegraphics[width=1.0\textwidth]{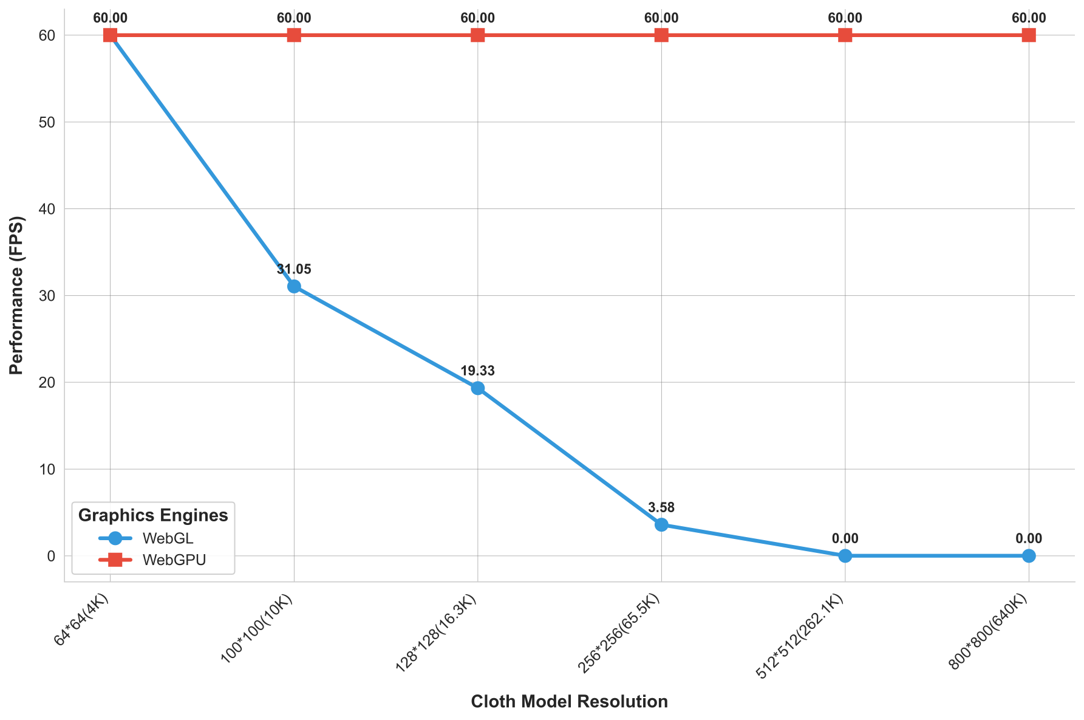}
	\caption{Performance comparison between WebGPU and WebGL in Hanging Simulation.}
	\label{fig:result1}
\end{figure}

The first scene is designed to evaluate the performance of WebGPU’s acceleration compared to WebGL. Specifically, we simulate a hanging cloth, where the suspended cloth model is influenced only by gravity and measure the performance difference between WebGPU and WebGL. Figure \ref{fig:result1} shows a performance comparison between WebGL and WebGPU.

This experiment measured  simulation performance  by updating only the cloth model while excluding collision processing algorithm of the cloth object and other objects. , The experiment started with 4K cloth nodes in both WebGPU and WebGL environments, gradually increasing the cloth resolution to determine the maximum achievable real-time resolution. At this point, the resolution of the cloth object that satisfies 30fps and the cloth object resolution that can be simulated at the maximum were found. The experiment confirmed that the maximum allocation limit resolution in WebGPU was 640K nodes, which is the limit of 128MB of memory, which is the buffer allocation of the WebGPU. In WebGPU, parallel-based acceleration using GPU was possible, so 60fps could always be ensured without problems at any resolution. On the other hand, in WebGL, parallel-based acceleration using GPU was not possible, so it was confirmed that real-time simulation was difficult of 10K or more cloth nodes. In particular, at the resolution of 65.5K cloth nodes, it was found that normal simulation was impossible at a performance of 3.58fps. It can be seen that in order to satisfy the same 60fps simulation performance in both environments, it is necessary to use a cloth object with a resolution of about 160 times smaller.

\subsection{Observation the Performance Limitations of WebGPU: Cloth Model Collide with 3D Surface Models}
\begin{figure}[H]
	\centering
	\includegraphics[width=1.0\textwidth]{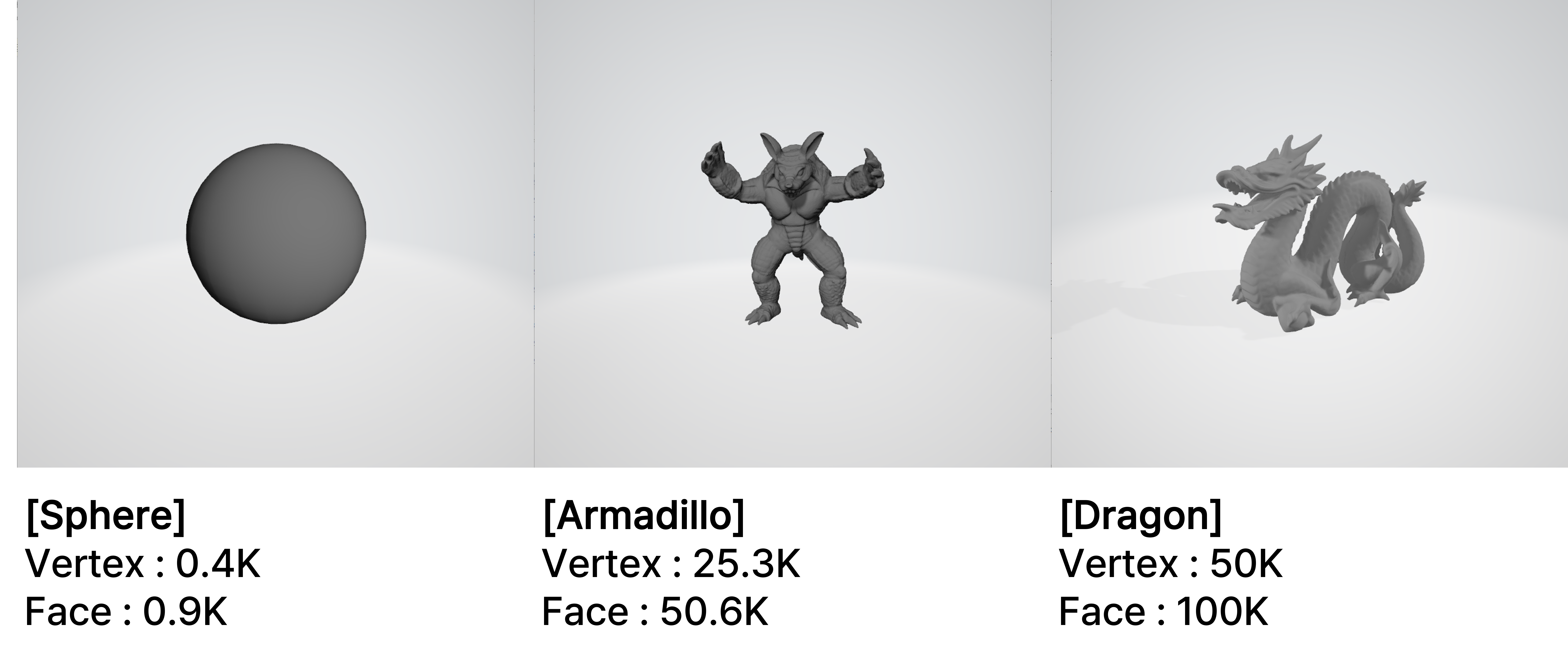}
	\caption{Information about the 3D surface models used in the second experiment.}    
	\label{fig:3d model info}
\end{figure}

The second experiment examines the limitations of WebGPU’s acceleration performance, particularly regarding collision processing and response with 3D surface model. In the second experiment, three types of 3D surface models were baked and the cloth models were collided, and the collision response that did not penetrate the object by pulling the fabric in one direction was confirmed. Figure~\ref{fig:3d model info} presents the information of the 3D models used in the second experiment. The Sphere model consists of 0.4K verticies and 0.9K triangles. The Armadillo model consists of 25.3K verticies and 50.6K triangles. Finally, the Dragon model consists of 50K vertices and 100K triangles.

\begin{figure}[H]
	\centering
	\includegraphics[width=1.0\textwidth]{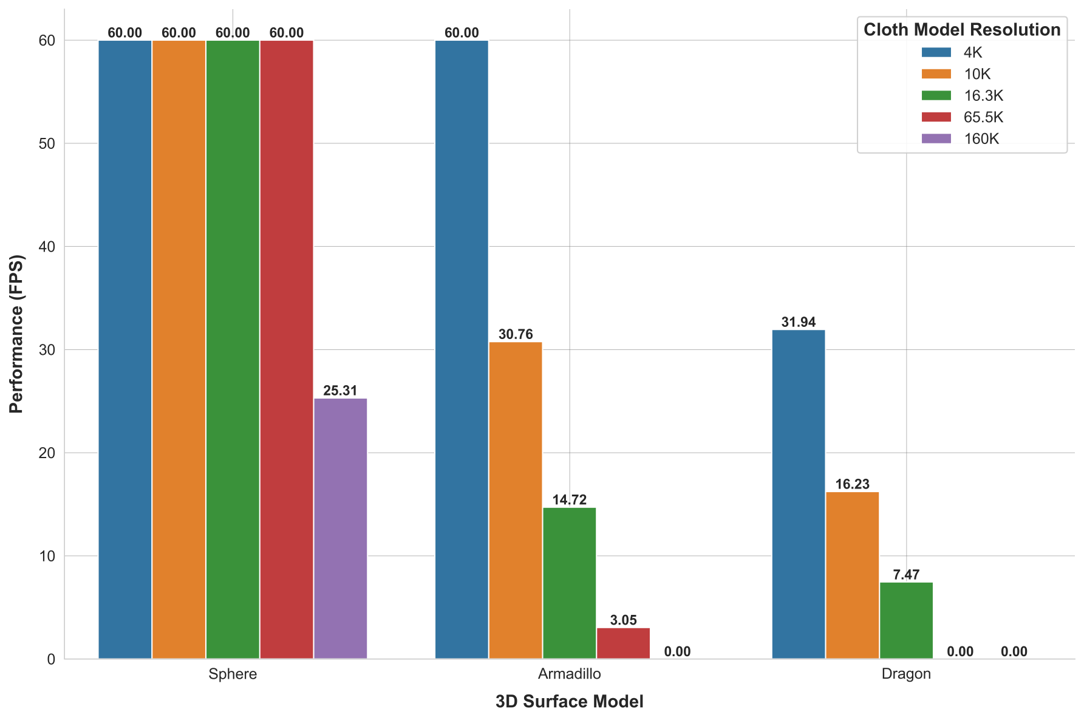}
	\caption{Experimental Results: Performance Comparison with Changing Resolution of Model}    
	\label{fig:result2}
\end{figure}

We conducted a performance comparison by increasing the resolution of the cloth model for the three models introduced in Figure~\ref{fig:result2}. The designed limitation indicate that 30fps cannot be achieved. Therefore, we explored cases where 30fps could not be achieved by incrementally increasing the resolution.

As a result of the experiment, it was confirmed that 30fps was achievable when simulating a collision between the Armadillo model with 50,000 triangles and a cloth object with a 10K cloth node resolution. At this time, the maximum resolution of the cloth object that could be simulated was 65.5K, at which the performance dropped to 3.05fps, far below real-time requirements, making the simulation nearly impossible. If the cloth model resolution exceeded this threshold, this was generated, the simulation performance degraded to the extent that frames per second could not be measured. Next, when simulating the collision between the Dragon model with 100,000 triangles more complex than the Armadillo model and the cloth object with a 4K cloth node resolution, we confirmed that 30fps was achievable. The maximum resolution that could be simulated was 16.3K, and the performance during this time was 7.47fps, which was about half the performance compared to the Armadillo model.

Since the accuracy of collision operations with other objects decreases as the cloth model resolution(number of cloth nodes) and 3d model resoltion(number of vertices and faces) is reduced, the highest possible resolution should be used to obtain realistic and appropriate simulation results. However, using a high-resolution model often makes real-time rendering impossible due to the increased computational load, so selecting an appropriate balance between real-time performance and realistic rendering is necessary.

\subsection{Visualizing each Experiments}
\begin{figure}[H]
	\centering
	\includegraphics[width=0.85\textwidth]{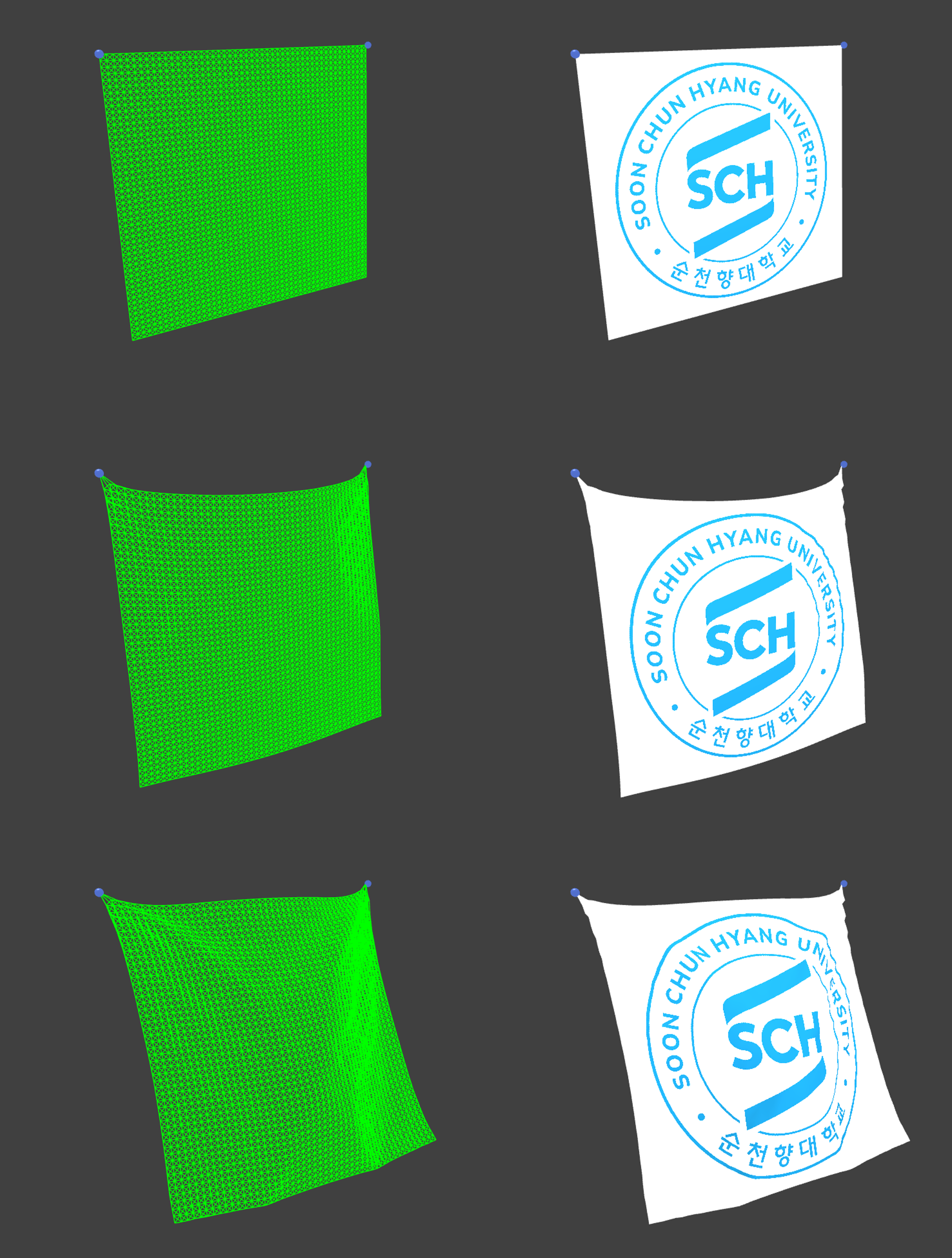}
	\caption{Scene1. Haning cloth simulation results(left: wireframe rendering, right: texture rendering)}    
	\label{fig:ex1}
\end{figure}

\begin{figure}[H]
	\centering
	\includegraphics[width=1.0\textwidth]{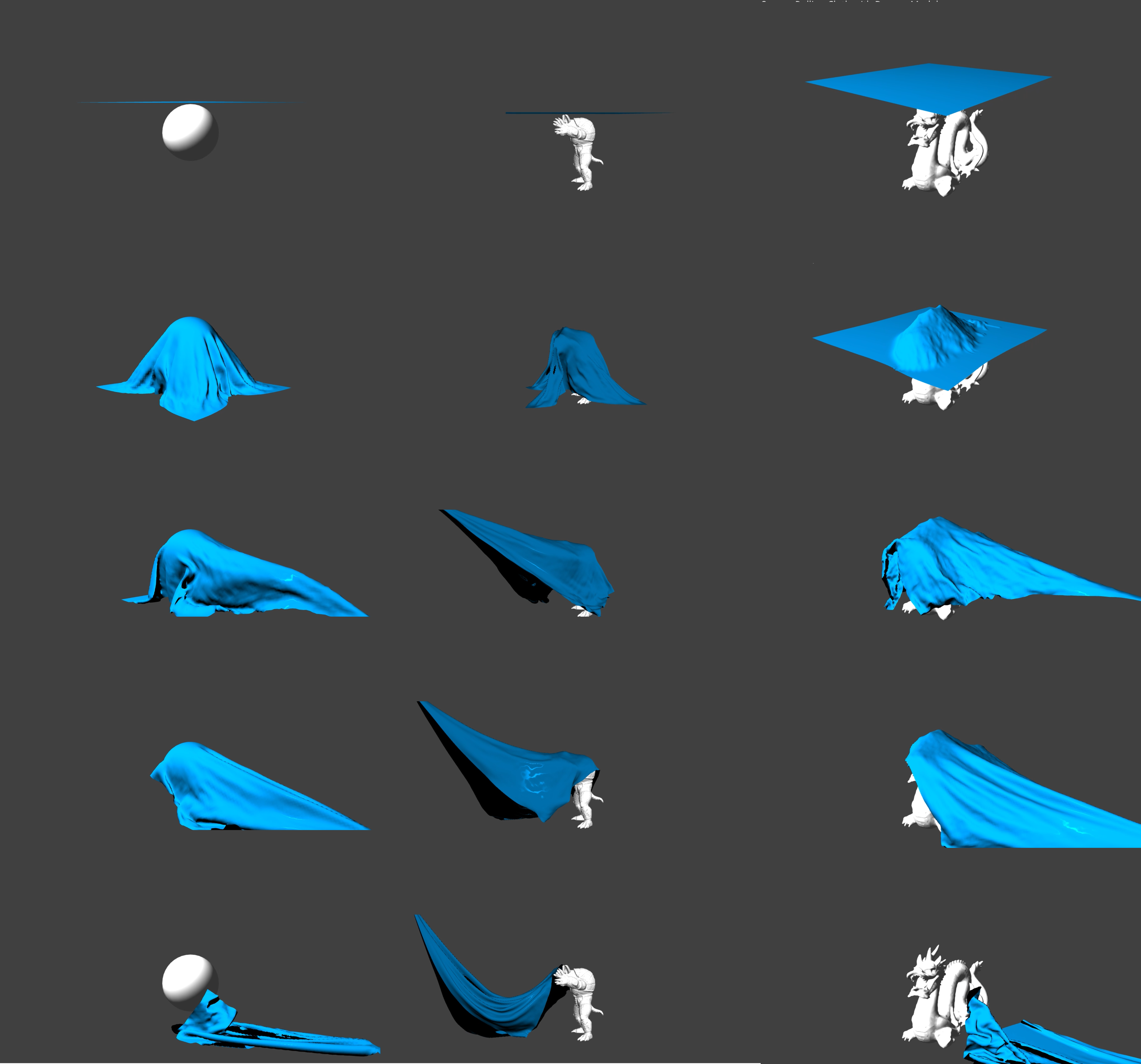}
	\caption{Scene2. Cloth-puling simulation results with 3D objects.(left: sphere model interaction, middle: armadillo model interaction, right: dragon model interaction)}	
	\label{fig:ex2}
\end{figure}

Figure \ref{fig:ex1} and \ref{fig:ex2} represent the simulation results of each experiment. 


%% file: conclusion.tex
\section{Discussion}
WebGPU, introduced in 2024, significantly improves the feasibility of GPU acceleration in web-based graphics environments, where GPGPU was  previously impractical. In our experiments, we explored the usability of WebGPU as a new web graphics paradigm by comparing its performance with WebGL and evaluating its performance limits. The results clearly show that, for high-resolution cloth simulation, WebGPU significantly outperforms WebGL. Specifically, WebGPU maintained 60fps with up to 640K nodes, while WebGL struggled to deliver real-time performance beyond 10K nodes. This performance boost is attributed to WebGPU’s efficient utilization of GPU parallelism, making it highly suitable for computationally intensive tasks such as cloth simulation and complex collision handling. Our contributions include the development of a detailed collision detection and response system based on an extended M¨oller–Trumbore algorithm, as well as a spring-centric force calculation method that enables real-time simulation even in  high- complexity scenarios. Furthermore, our experiments using a mass-spring system to manage collisions between a cloth model and a 3D model revealed that even at increased resolution—such as a cloth with 4K nodes colliding with a 100K triangle Dragon model—WebGPU sustained 30fps.  These findings not only validate the potential of WebGPU for real-time simulations but also highlight the need for further optimizations, such as implementing barycentric-coordinate optimization~\cite{sung2018optimization}, collision culling techniques~\cite{govindaraju2004fast,liu2010real,zhang2007continuous,barbic2010subspace}, and alternative collision detection algorithms like the GJK algorithm~\cite{gilbert1988fast,montaut2024gjk}. These enhancements will help balance performance with realistic rendering.

\section{Conclusion}
In summary, our study confirms the viability and advantages of WebGPU for real- time simulation in web-based graphics applications. The significant performance improvements over WebGL, especially in high-resolution simulations and complex collision-handling scenarios, indicate that WebGPU is a promising technology for future applications in gaming, animation, and VR/AR/XR content. Nevertheless, further improvements are necessary, particularly in reducing the computational overhead of per-frame collision checks and extending the simulation framework to volumetric objects. Future work will involve integrating advanced solvers such as Position-based Dynamics (PBD), Projective Dynamics (PD), and Vertex Block Descent (VBD) Additionally we plan to adopt per-formance optimizations like bounding volume hierarchy (BVH) for collision culling. We believe that these enhancements will further advance web-based graphics technologies and broaden the applicability of our simulation system.

%% file: appendix.tex
\section{Appendix}
\subsection{Spring-centric Cltoh Simulation Algorithm}
Algorithm~\ref{alg:Spring-centric Force Calculation Algorithm}, computes the forces exerted by each spring connecting two nodes based on Hooke’s law. For each spring (processed into a workgroup thread), it retrieves the indices of the connected nodes and calculates the normalized direction between them. The spring force is computed as the product of the stiffness coefficient and the deviation from the spring’s rest length, while the damping force is computed using the difference in the nodes velocities along that direction. The estimated force is then the sum of these two components, applied in opposite directions to the two nodes (according to Newton’s third law) via atomic additions to a temporary force buffer.

\begin{algorithm}[H]
	\caption{Spring-centric Force Calculation Algorithm}
    \label{alg:Spring-centric Force Calculation Algorithm}
	\begin{algorithmic}[1]
		\State \textbf{Input:} global\_id: vector3\_uint
		\State \textbf{Output:} Calculated Temporary Force Buffer
		\State \textbf{Procedure}: UpdateSpringForce (global\_id)		
		\State $\text{@workgroupSize} \text{ (256, 1, 1)}$
		\State $index \gets global\_id.x$
		\State $\text{Buffer Force} \gets \text{Node Force Buffer Vector3[numNodes]}$
		\State $node1 \gets \text{Springs}[index].indexA$
		\State $node2 \gets \text{Springs}[index].indexB$		
		\State
		\State $\text{Calculate Spring Forces Part}$\Comment{Hooke's law}
		\State $dir \gets \Call{normalize}{node2.pos - node1.pos}$
		\State $springForce \gets stiffness \times (\Call{Length}{pos2 - pos1} - restLength$
		\State $dampingForce \gets damping \times (\Call{Dot}{node2.vel, dir} - \Call{Dot}{node1.vel, dir})$
		\State $estimatedForce \gets (springForce + dampingForce) \times dir$
		\State
		\State $\text{Vector3 force} \gets \text{Spring Force Equation Result for Temporary}$
		\State $\text{AtomicAdd}(\text{Force}[node1], \text{estimatedForce})$ \Comment{The law of Action-Reaction}
		\State $\text{AtomicAdd}(\text{Force}[node2], -\text{estimatedForce})$ \Comment{The law of Action-Reaction}
		\State $\textbf{End Procedure}$
	\end{algorithmic}
\end{algorithm}

\clearpage
\subsection{Collision Response Algorithm}
Algorithm~\ref{alg:Collision Response and Position Update Algorithm}, each thread of the workgroup processes a single cloth vertex identified by its global index. First, the algorithm retrieves the previous position and current velocity of the vertex. It then reads the cumulative collision response direction and the corresponding count from the temporary atomic buffers. If the count is greater than zero, indicating that one or more collisions have been detected for the vertex, the velocity is reversed and dampened, and the vertex position is updated by adding the computed response direction. The updated velocities and positions are then written back to the appropriate arrays. Finally, the temporary buffers are reset using the atomic store operations to prepare for the next simulation step.
\begin{algorithm}
	\caption{Collision Response and Position Update Algorithm}
    \label{alg:Collision Response and Position Update Algorithm}
	\begin{algorithmic}[1]
		\State \textbf{Input:} global\_id: vector3\_uint
		\State \textbf{Output:} Updated velocities and positions for cloth vertices
		\State \textbf{Procedure}: Apply Collision Response (global\_id)
		\State $\text{@workgroupSize} \text{ (256, 1, 1)}$
		\State $index \gets global\_id.x$
		\If{$index > \text{numParticles}$}
		\State \textbf{return}
		\EndIf		
		\State $pos \gets \text{getPrevPosition}(x)$ \Comment{Get previous position}
		\State $vel \gets \text{getClothVertexVelocity}(x)$ \Comment{Get velocity}
		\State $responseDirection \gets \text{atomicLoad}(\&\text{tempBuffer}[x].value)$
		\State $countBufferData \gets \text{atomicLoad}(\&\text{tempCountBuffer}[x].value)$
		\If{$countBufferData > 0$}
		\State $vel \gets vel \times -0.5$ \Comment{Invert and reduce velocity}
		\State $\text{pos} \gets pos + responseDirection$
		
		\State $\text{velocities}[x \times 3 + 0] \gets vel.x$
		\State $\text{velocities}[x \times 3 + 1] \gets vel.y$
		\State $\text{velocities}[x \times 3 + 2] \gets vel.z$
		
		\State $\text{positionsCloth}[x \times 3 + 0] \gets pos.x$
		\State $\text{positionsCloth}[x \times 3 + 1] \gets pos.y$
		\State $\text{positionsCloth}[x \times 3 + 2] \gets pos.z$
		
		\EndIf
		
		\State $\text{atomicStore}(\&\text{tempCountBuffer}[x].value, \text{i32}(0))$ \Comment{Reset count buffer}
		\State $\text{atomicStore}(\&\text{tempBuffer}[x].value, \text{i32}(0))$ \Comment{Reset temp response direction buffer}
		\State $\textbf{End Procedure}$
	\end{algorithmic}
\end{algorithm}